\documentclass[aps,prb,superscriptaddress,twocolumn]{revtex4}
\usepackage{amsmath,amssymb,bm}
\usepackage{graphicx}
\usepackage{epstopdf}
\usepackage{latexsym}
\usepackage{subfigure}
\usepackage{color}
\usepackage{amssymb}
\usepackage{wasysym}
\usepackage{verbatim}

\newcommand{\la}{\langle}
\newcommand{\lr}{\rangle}

\begin{document}
\date{\today}
\title{Finite size effects in the $Z_2$ spin liquid on the kagome lattice}

\author{Hyejin Ju}
\affiliation{Department of Physics, University of California, Santa Barbara, Santa Barbara, CA, 93106}

\author{Leon Balents}
\affiliation{Kavli Institute of Theoretical Physics, University of California, Santa Barbara, Santa Barbara, CA, 93106}

\begin{abstract}
  Motivated by the recent discovery of the $Z_2$ quantum spin liquid
  state in the nearest neighbor Heisenberg model on the kagome
  lattice, we investigate the ``even-odd" effect occuring when this
  state is confined to infinitely long cylinders of finite
  circumference.  We pursue a dual analysis, where we map the
  effective $Z_2$ gauge theory from the kagome lattice to a frustrated
  Ising model on the dice lattice.  Unexpectedly, we find that the
  latter theory, if restricted to nearest neighbor interactions, is
  insufficient to capture this effect.  We provide an explanation of
  why further neighbor interactions are needed via a high-temperature
  expansion of the effective Hamiltonian.  We then carry out 
  projective symmetry group analysis to understand which second
  neighbor interactions can be introduced while respecting the lattice
  symmetries.  Finally, we qualitatively compare our results to
  numerics by computing the dimerization operator within our theory.
  Systems with odd circumferences exhibit a non-vanishing dimerization
  that decays exponentially with circumference.
\end{abstract}

\maketitle

\section{Introduction}

The kagome lattice with nearest neighbor antiferromagnetically coupled spins is one of the most frustrated systems in two dimensions.
A spin-1/2 system on this model has been one of the most promising candidates to realize a quantum spin (QSL) liquid ground state, where there is no symmetry breaking at zero temperature due to quantum fluctuations and geometric frustration~\cite{balents2010spin}.
A long standing goal has been to identity an experimental realization of this system, and huge efforts have been made to not only grow such materials, but also to understand their ground states.
For instance, single crystals of Herbertsmithite~\cite{matthew2005structurally,helton2007spin,han2012refining}, which is composed of kagome planes with $\text{Cu}^{2+}$ as its magnetic centers, have been successfully engineered, and since then, there have been numerous experiments from neutron scattering~\cite{han2012refining} to NMR to check for a spin liquid ground state.

From the theoretical end, a plethora of papers have numerically investigated the nearest neighbor Heisenberg model on the kagome lattice.
Earlier studies, based on dimer model approaches as well as series expansions, have suggested that the ground state is {\sl not} a spin liquid, but a valence bond solid~\cite{marston1991spin,nikolic2003physics,singh2007ground,jiang2008density,evenbly2010frustrated}, a state that breaks translational symmetry while respecting spin rotation symmetries.
However, a recent numerical study~\cite{yan2011spin} has unveiled a possibility that the ground state of this model harbors an exotic $Z_2$ spin liquid state~\cite{read1991large,PhysRevB.44.2664,PhysRevB.45.12377,wang2009z_,tay2011variational,lu2011z_,iqbal2011valence,iqbal2011projected}.
In this work, done by using Density Matrix Renormalization Group (DMRG) methods~\cite{white1992density,white1993density}, it has been shown that finite spin gaps in both the singlet and triplet sectors exist and that spin correlations are uniform throughout the lattice.
In addition, this work has shown that there may be some spontaneously translational symmetry breaking, which is deeply rooted in the Lieb-Schultz-Mattis theorem~\cite{lieb1961two}, in cylinders with an odd circumference.
Furthermore, another numerical study~\cite{jiang2012identifying} has further corroborated this claim via the measurement of the topological entanglement entropy, a universal quantity which has a specific non-zero value of $\ln 2$ for ground states with non-trivial topology~\cite{levin2006detecting,kitaev2006topological}.
Through the combinational efforts of these two numerical studies, it has been widely accepted that the ground state of the nearest neighbor Heisenberg model on the kagome lattice indeed fosters the exotic $Z_2$ spin liquid.

In this paper, we explore the effects of confining this model onto infinite length cylinders with finite circumferences.
Specifically, we consider the effective $Z_2$ gauge theory on the kagome lattice and predict that these systems with odd circumferences have a non-vanishing dimerization.
This ``even-odd" effect was first considered in Ref.~\onlinecite{yao2012exact} for quantum dimer models and then in Ref.~\onlinecite{jiang2012spin} on the analysis of the $J_1-J_2$ Heisenberg model on the square lattice.
In order to understand the origin of the even-odd effect, it is instructive to discuss the effective $Z_2$ gauge theory on the square lattice and adapt the following argument from Ref.~\onlinecite{jiang2012spin}.
We claim that this finite-size effect is not a distraction from the
physics, but rather an intrinsic property of the $Z_2$ QSL phase.  

Consider the following $Z_2$ gauge theory description of the QSL~\cite{senthil2000z_}, given by the following Hamiltonian
\begin{equation}
  \label{eq:hamikagome}
  H = -K \sum_{\text{plaquette}} \prod_{\langle i j\rangle \in \text{plaq}} \sigma_{ij}^z - h \sum_{\langle ij \rangle} \sigma_{ij}^x,
\end{equation}
and
\begin{equation}
  \label{eq:oddkagome}
  \prod_j \sigma_{ij}^x = -1.
\end{equation}
Here, the sum is over all plaquettes of the square lattice.
These two equations describe the ``odd Ising gauge theory", where it describes the deconfined quantum spin liquid for $K \gg h$ and an ordered state for $K \ll h$.
Ref.~\onlinecite{jiang2012spin} computes the dimerization operator and show that this is the quantity that exhibits the even-odd effect.
Additionally, one can consider the following operator
\begin{equation}
Q_x = \prod_{y = 1}^{L_y} \sigma_{xy;x+1,y}^x
\end{equation}
which is a string of $\sigma^x$s, centered around $x,x+1$ that wrap around the cylinder.
While this is not exactly the dimerization operator, one can heuristically argue that this operator already breaks translational symmetry for systems with an odd circumference.
Here, $Q_x$ commutes with the Hamiltonian, Eq.~\eqref{eq:hamikagome}, and thus, is a constant of motion.
Imagine, for instance, a system with two open ends.
Then, with the aid of Eq.~\eqref{eq:oddkagome}, we can rewrite $Q_x$ as
\begin{equation}
Q_x = \prod_{y=1}^{L_y} \left( \prod_{|j-i|=1} \sigma_{ij}^x \right)_{i = (x,y)} = (-1)^{x L_y},
\end{equation}
which can be derived inductively by considering the open end at $x=1$
first.  From this operator, we can see that $Q_x = 1$ always for an even
$L_y$; however, $Q_x$ oscillates along the chain for odd $L_y$.  This
indicates that the $Z_2$ gauge theory ``knows'' globally that
translational symmetry is broken for odd $L_y$, even though all local
gauge-invariant observables appear na\"ively translationally symmetric
(and indeed are in the two-dimensional (2d) thermodynamic limit, in the deconfined phase).  

We take a similar approach to study the $Z_2$ gauge theory on the kagome lattice.
The main difference between the work in Ref.~\onlinecite{jiang2012spin} and our work is that the nearest neighbor model, as above on the square lattice, is not sufficient to capture the even-odd effect on the kagome lattice.
This is much clearer to see in the dual Hamiltonian on the dice lattice, where we employ both high-temperature expansions and projective symmetry group (PSG) analysis~\cite{PhysRevB.65.165113} in Section.~\ref{sec:eo}.
The need for further neighbor interactions arises from the complicated structure of the dual lattice, which for this case is the dice lattice that consists of three independent sublattices in a hexagonal array.
We give a heuristic argument in Section~\ref{sec:hight} to give the reader an intuition as to why such an analysis is needed. 

The remainder of this paper is organized as follows.
In Section~\ref{sec:model}, we introduce the effective $Z_2$ gauge theory on the kagome lattice and map this model unto its dual equivalent on the dice lattice.
We discuss the relevant symmetries and introduce our gauge choices on the dice lattice.
In Section~\ref{sec:eo}, we first use high-temperature expansions to show that further neighbor interactions are required to obtaining the even-odd effect.
Furthermore, we explore this state in detail to compute the PSG of this model to include symmetry-allowed second neighbor interactions.
We also present our main results in this section and finally conclude in Section~\ref{sec:conclude}.
The Appendices include many derivations and calculations, such as detailed PSG analysis needed for the second half of the paper.

\section{Model and notation}
\label{sec:model}

\subsection{Ising gauge theory}
\label{sec:ising-gauge-theory}

Motivated by the analysis in Ref.~\onlinecite{jiang2012spin}, we discuss here an effective $Z_2$ gauge theory description of the QSL state on the kagome lattice.
Because we are mainly focused on the behavior of the dimerization of this state, we can integrate out the spinons, as Ref.~\onlinecite{yan2011spin} discussed that a finite spin gap exists on this state.
In doing so, we can reduce the Hamiltonian to the following
\begin{equation}
  \label{eq:hamikagome}
  H = -K \sum_{\text{plaquette}} \prod_{\langle i j\rangle \in \text{plaq}} \sigma_{ij}^z - h \sum_{\langle ij \rangle} \sigma_{ij}^x,
\end{equation}
where the $\sigma_{ij}$ live on the bonds $(ij)$ of the kagome lattice.
This description is of the ``odd" Ising gauge theory, where 
\begin{equation}
  \label{eq:oddkagome}
  \prod_j \sigma_{ij}^x = -1.
\end{equation}
For $K \gg h$, the system is in a deconfined QSL state, while for $h \gg K$, it is in an ordered, confined state.
We can consider the dimerization operator to understand the translational symmetry breaking of this state as it is confined to a finite circumference cylinder.
We will commonly refer to this as the ``even-odd" effect, as there will be spontaneous symmetry breaking for an odd cylinder and not for an even one.
On symmetry grounds, we expect that $\langle D_i^x \rangle \propto \langle \sigma_{ij}^x \rangle$.
The goal is to compute this expectation value and show that it is exactly this operator that breaks translational symmetry along the length of the cylinder. 

%
\begin{figure}[t]
  \begin{center}
  \scalebox{0.5}{\includegraphics[width=\columnwidth]{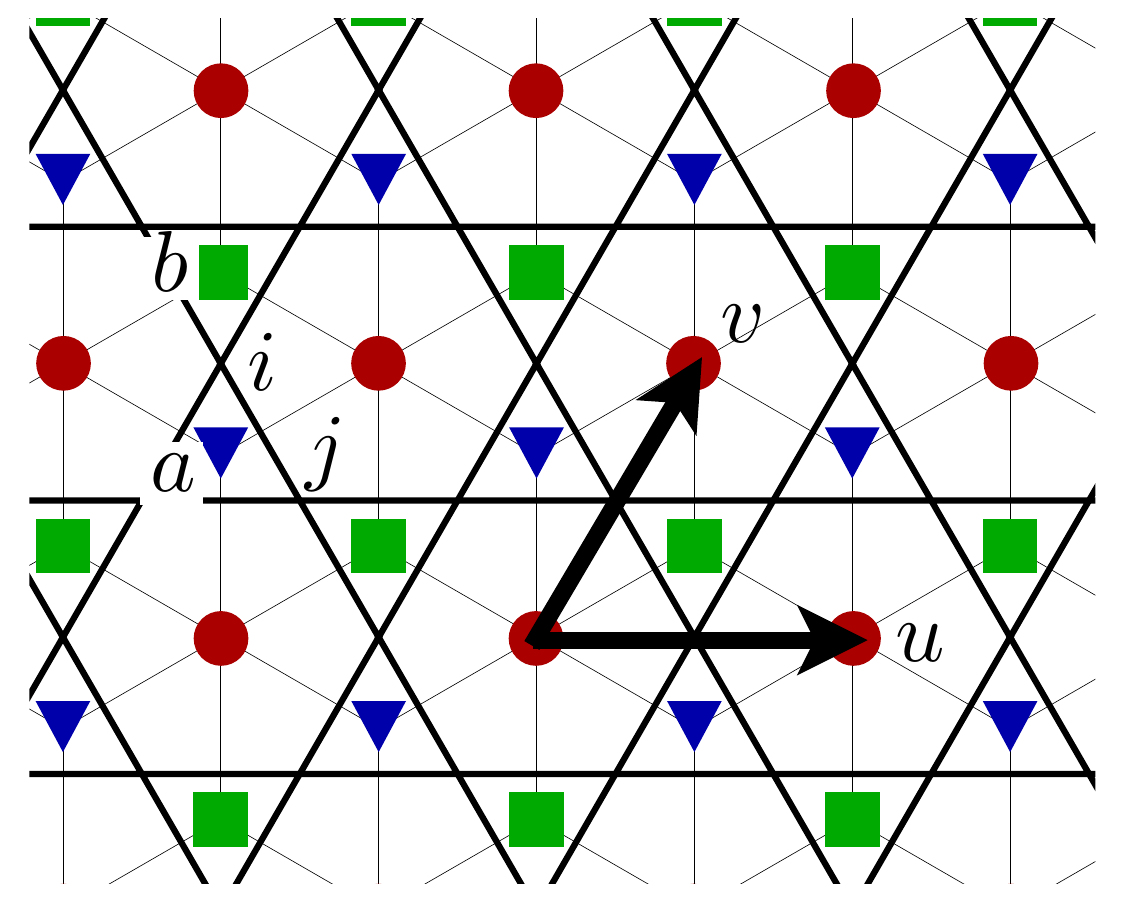}}
  \end{center}
  \caption{The kagome lattice (thick line) with its dual, dice lattice (thin line).
  		The kagome sites are labeled by $i,j$ and the dice sites by $a,b$.
		The dice lattice has three independent sites, shown as circles (red), squares (green), and triangles (blue), and has translational symmetries under $\mathbf{u}$ and $\mathbf{v}$, as shown.
		It also has a 6-fold rotational symmetry about the (red) circular site, as well as a 3-fold reflection symmetry about the three axes crossing the (red) circular site.}
        \label{fig:duality}
\end{figure}
%

\subsection{Dual transverse field Ising model}
\label{sec:dual-transv-field}

To do so, we take a duality mapping, which provides a more intuitive understanding of this effect.
The duality transformations of Eqs. (\ref{eq:hamikagome},\ref{eq:oddkagome}) is done with the following
\begin{eqnarray}
  \label{eq:dual1}
   \tau_a^x & = & \prod_{\langle ij \rangle \in \text{plaq.} a} \; \sigma_{ij}^z\\
   \label{eq:dual2}
   \sigma_{ij}^x & = & J_{ab} \tau_a^z \tau_b^z.
\end{eqnarray}
Here, the site labels $i,j$ denote sites on the kagome lattice while $a,b$ denote sites on the dice lattice, as shown in Fig.~\ref{fig:duality}.
The $\langle ij \rangle$ are bonds associated with the dual site $a$ (e.g. the up triangle for the triangular (blue) site on the dice lattice).
Dual sites $a,b$ are the bonds on the dice lattice that intersects perpendicularly with a kagome bond $\la ij \lr$.
The factor $J_{ab}$ is introduced and must be chosen to satisfy Eq.~\eqref{eq:oddkagome}, which requires that its product around the dual plaquette, or parallelogram, is equal to -1, i.e.
\begin{equation}
  \label{eq:odddice}
  \prod_{\text{parallelogram}}\; J_{ab} = -1.
\end{equation}
Then, the dual odd Ising gauge Hamiltonian can be written as follows
\begin{equation}
	\label{eq:dicehami}
	H = -h \sum_{\langle ab \rangle} J_{ab} \tau_a^z \tau_b^z - K \sum_a \tau_a^x.
\end{equation}
Notice that model is invariant under $Z_2$ transformations, namely,
\begin{eqnarray}
	\label{eq:3b}
	J_{ab} & \to & s_a s_b J_{ab} \\
	\tau_a^z & \to & s_a \tau_a^z, \nonumber
\end{eqnarray}
where $s_a = \pm 1$.

\subsection{Symmetries}
\label{sec:symmetries}

Before we dive into our gauge choices for $J_{ab}$, we first comment on the symmetries of the dice lattice.
The dice lattice has three independent sites per hexagonal unit cell, shown in Fig.~\ref{fig:duality} as the 6-coordinated circular (red) site and 3-coordinated square (green) and triangular (blue) sites.
It is invariant under hexagonal translations, $T_u, T_v$ as well as point symmetries, which include $\pi/3$ rotations, $R_{\pi/3}$, and three-fold reflections about the circular (red) site.
These reflection symmetries are not independent, however, and we only need to consider {\sl one} of the three reflections.
We give further details in Appendix~\ref{ap:equiv}.

%
\begin{figure}[t]
  \begin{center}
  \scalebox{0.7}{\includegraphics[width=\columnwidth]{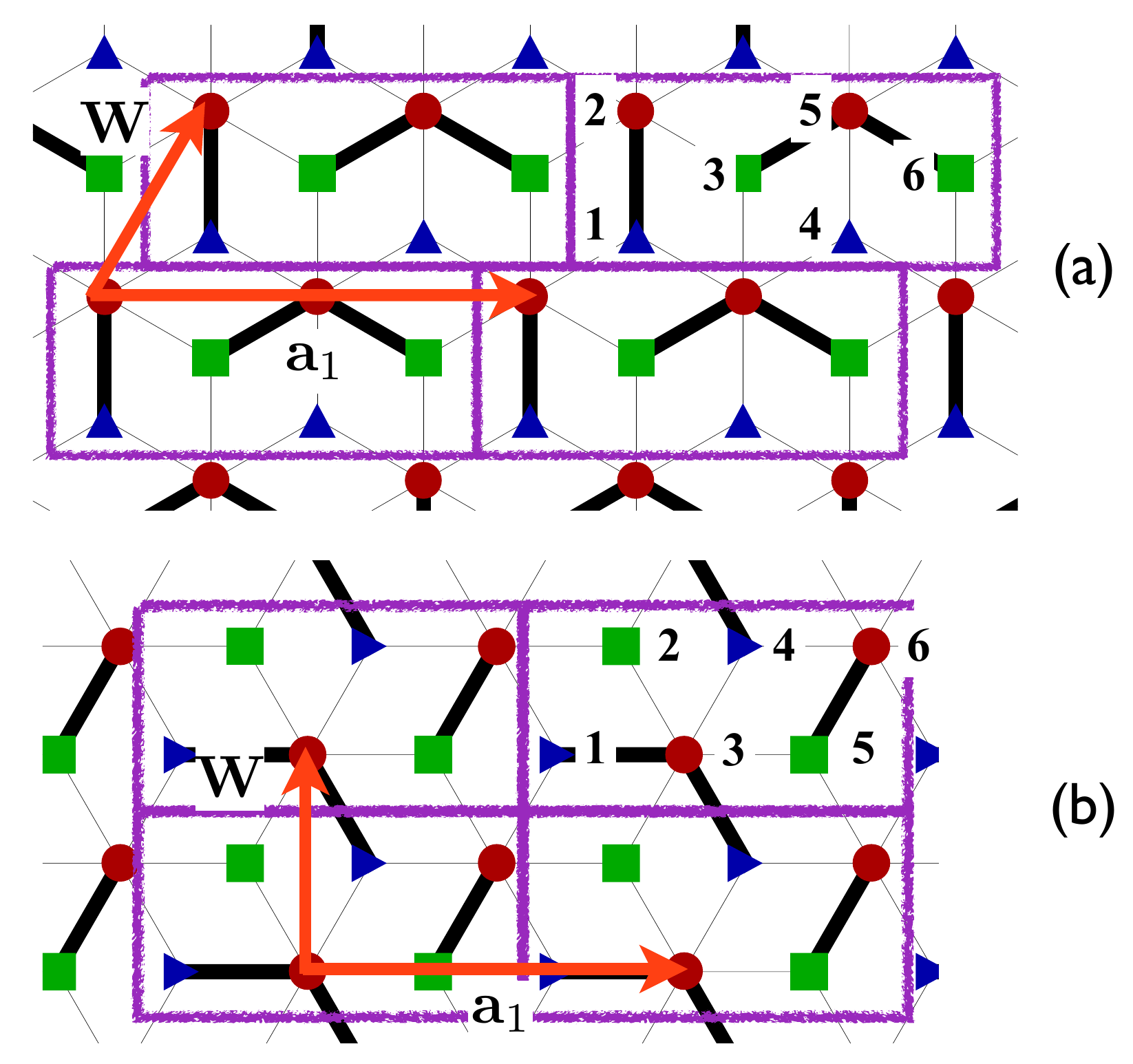}}
  \end{center}
  \caption{Gauge choices for (a) twisted and (b) straight boundary conditions. Here, $\mathbf{W}$ is the winding vector and $\mathbf{a}_1$ is the infinite direction.
  		The thick lines depict where $J_{ab} = -1$ while the thin lines show $J_{ab} = 1$.
		Note that the lattice orientation of (a) is at $90^o$ from that of (b).}
        \label{fig:gauge}
\end{figure}
%

\subsection{Gauge choices}
\label{sec:gauge-choices}

In order to proceed, we need to choose a configuration of $J_{ab}$
that satisfies Eq.~\eqref{eq:odddice}.  To treat 
quasi-one-dimensional (1d) cylinders, we need a gauge choice that is
as 1d as possible.  A recent study of the 2d version of
this model~\cite{huh2011vison} investigated the vison confinement
transitions between the $Z_2$ gauge theory and valence bond solid
states.  There, they considered a magnetic unit cell that encompasses
{\sl two} legs of the kagome lattice, which is unable to capture the
even-odd effect.  In this paper, we consider two separate
one-dimensional gauge choices, one with ``twisted" and the other with
``straight" boundary conditions.  (In notations used in
Ref.~\onlinecite{yan2011spin}, this corresponds to XC-2 and YC
geometries, respectively.)   We show our gauge choices in
Fig.~\ref{fig:gauge}, where the lattice is infinite along
$\mathbf{a}_1$ and winds along the winding vector, $\mathbf{W}$, for
both boundary conditions.  The labels $1, ..., 6$ denote the
sublattices within each magnetic unit cells.  The thick black lines
show bonds where $J_{ab} = -1$ while the thin lines show bonds with
$J_{ab} = 1$.  These gauge choices satisfy the constraint that arises
from the odd Ising gauge theory, Eq.~\eqref{eq:odddice}, where the
product of $J_{ab}$ around each plaquette is equal to -1.  In
Fig.~\ref{fig:gauge}(a), we show the ``twisted" boundary conditions,
where the cylinder has circumference $\frac{3}{2} L_y$ (measuring
parallel to the Cartesian $y$-axis), where $L_y$ is the number of unit
cells stacked along $\mathbf{W}$.  On the other hand,
Fig.~\ref{fig:gauge}(b) shows the ``straight" boundary conditions,
with cylinders of circumference $\sqrt{3} L_y$.  We will refer to our
gauge choices by the same terms used for the boundary conditions,
i.e. we will speak of ``twisted'' and ``straight'' gauge choices.

\section{Even and odd effect}
\label{sec:eo}

In this section, we use tools developed in the previous section to study
the even/odd effect.  Given the two gauge choices in
Fig.~\ref{fig:gauge}, we start by considering the differences between
systems with even and odd circumferences to see how we can break the
translational symmetry of this model.  First, using high-temperature
expansion, we explain the reason why the nearest neighbor model is
insufficient and why further neighbor interactions are required to see
such an effect.  Then, we derive the PSG for the nearest neighbor model,
with further details provided in Appendices \ref{ap:psg1},\ref{ap:psg2},
to add in interactions that respects both the lattice and gauge
symmetry.  Finally, we compare our results to the DMRG results from
Ref.~\onlinecite{yan2011spin} by summing the spin-spin correlations on
each triangle of the kagome lattice.

\subsection{High-temperature expansion}
\label{sec:hight}

The dimerization expectation value of interest corresponds in the dual
picture to
\begin{equation}
\la J_{ab} \tau^z_a \tau^z_b \lr .
\end{equation}
We are interested in this quantity in the ``trivial'' phase of the
dual transverse field Ising model, and in how it exhibits
translational symmetry breaking in odd cylinders.  Though some
symmetry breaking is expected to be generic on the grounds of the
Lieb-Schultz-Mattis theorem, it turns out that the nearest neighbor
model of Eq.~\eqref{eq:dicehami} is non-generic in this regard for the
kagome lattice, and no symmetry breaking appears in the dimerization without further
neighbor coupling.  This is in contrast to a similar analysis in
Ref.~\onlinecite{jiang2012spin}.  There, the analysis was done on a
square lattice, where it is adequate to consider simply a nearest
neighbor model.  

Our goal in this subsection is to provide an intuitive understanding
to why the nearest-neighbor model fails to exhibit dimerization, and
how further neighbor coupling remedies this deficiency.  To do so, we
make a simplification of the model.  We are in principle interested
really in the quantum transverse field problem defined by
Eq.~\eqref{eq:dicehami}, and in particular, in the behavior when the
system in two dimensions at zero temperature is in the $Z_2$
deconfined phase, which corresponds to the situation $K \gg h$.  In
the Ising language, this is the ``trivial'' quantum paramagnetic (with
spins largely polarized along the $\tau^x$ direction) phase of the
Ising model.  Clearly in the Ising model, the trivial phase is
smoothly connected to the {\em high temperature} paramagnetic one.  We
expect all universal properties of the model to be qualitatively the
same in both regimes.  Thus, from the point of view of the dual
variables, we may as well replace the quantum fluctuations induced by
large $K$, which appears as the transverse field in
Eq.~\eqref{eq:dicehami}, with {\em thermal} fluctuations of classical
spins $\tau^z_a$.  This is convenient because strong thermal
fluctuations are technically relatively easy to treat via a high
temperature expansion, in comparison with the quantum treatment
necessary for the large transverse field limit.

Thus, we take a high-temperature expansion of the Hamiltonian 
\begin{equation}
	\label{eq:genhighThami}
	H = -h \sum_{ab} J_{ab} \tau_a^z \tau_b^z,
\end{equation}
where $\tau^z_a = \pm 1$.
In the subsequent discussions remaining in this subsection, we drop the superscript $z$ for simplicity.

To start, we write the partition function that corresponds to the Hamiltonian in Eq.~\eqref{eq:genhighThami}
\begin{equation}
\label{eq:parthight}
Z = \cosh(\beta h) \sum_{\tau_k = \pm 1} \; \prod_{\langle ab \rangle} \; \left[ 1 + x J_{ab} \tau_a \tau_b \right]
\end{equation}
where $x = \tanh(\beta h )$.
This is derived in Appendix~\ref{ap:hight}, where a quick refresher of the high-temperature expansion on the Ising model is outlined.
Notice that only {\sl closed loops} contribute to this partition function; open loop contributions vanish under the sum over $\tau_k = \pm 1$.
Each closed loop is weighted by the product of $xJ_{ab}$ for
all the links in the loop.  This product is proportional to the
``flux'' through the loop, and is clearly gauge invariant, and also
consequently translationally invariant, in the thermodynamic (2d)
limit.  
For instance, the smallest loop is any plaquette (or parallelogram) of
the dice lattice and contributes a factor of $-x^4$, since the product
of $J_{ab}$ on such a plaquette is equal to -1 in our theory.

Next, consider the following correlation function,
\begin{eqnarray}
	\label{eq:highTcorr}
	\displaystyle
	\langle J_{mn} \tau_m \tau_n \rangle  & = &  \frac{\sum_{\tau_k = \pm 1} \; \prod_{\langle ab \rangle} \; J_{mn} \tau_m \tau_n \left[ 1 + x J_{ab} \tau_a \tau_b \right]}
										{\sum_{\tau_k = \pm 1} \; \prod_{\langle ab \rangle} \; \left[ 1 + x J_{ab} \tau_a \tau_b \right]},\nonumber\\
\end{eqnarray}
where we normalize, as usual, by the partition function given in Eq.~\eqref{eq:parthight}.
This correlation function can be obtained order by order in $x$ in the
high-temperature expansion.  The expansion of the numerator also takes
the form of loops, and differs from that of the partition function by
the fact that the loops in the numerator are restricted to contain the
specific bond $(mn)$ which is being measured.  We see that the results
are clearly gauge and translationally invariant in 2d.

\begin{figure}[t]
	\begin{center}
	\scalebox{0.32}{\includegraphics[]{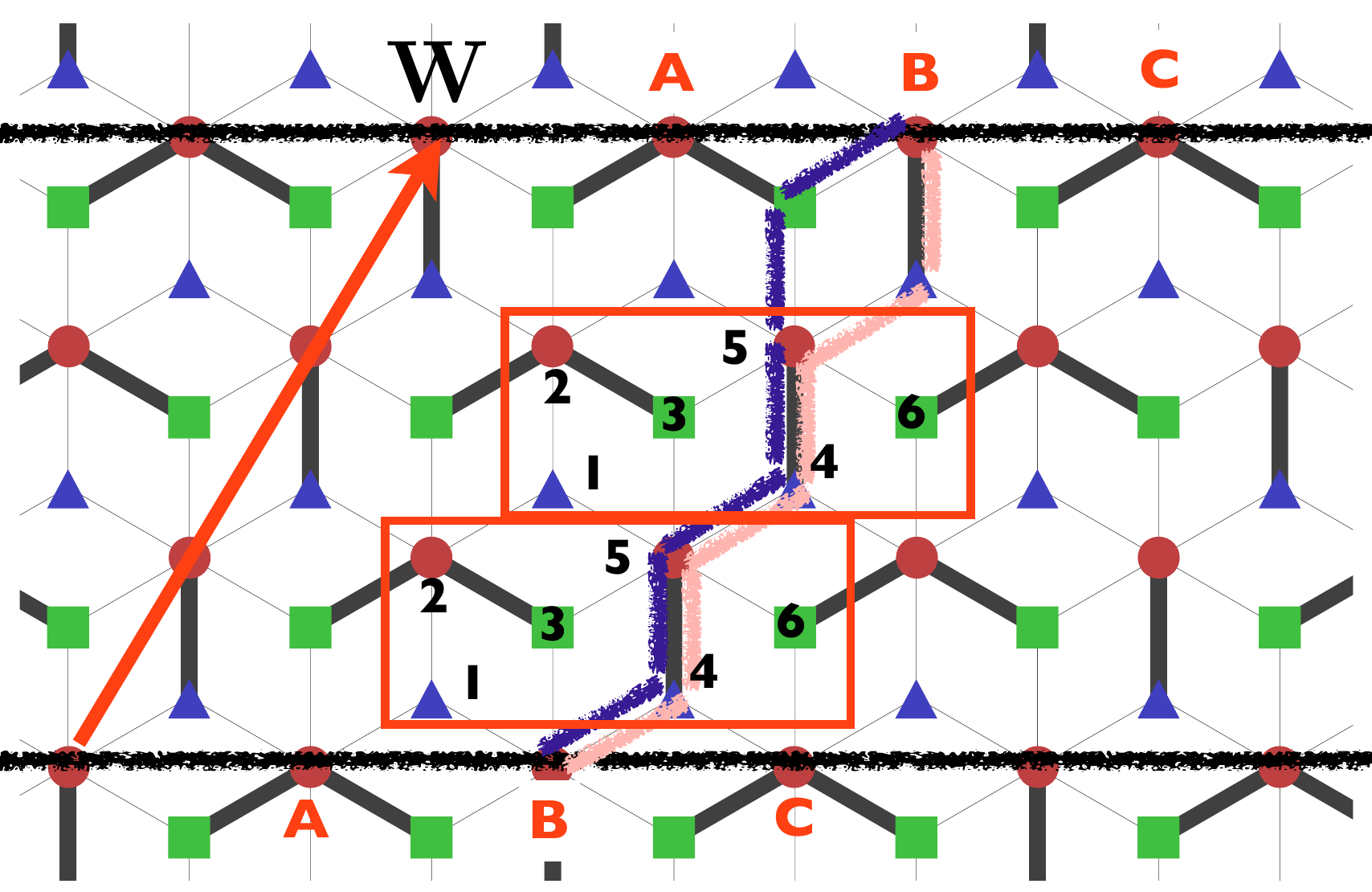}}
	\end{center}
	\caption{Local paths around the cylinder for the nearest-neighbor model always come in pairs (shown in dark (blue) and light (red) paths), which have the same value with different signs. Therefore, we expect that the ratio nearest neighbor correlations will always be equal 1 (e.g. $\langle \tau_1 \tau_2 \rangle/ \langle \tau_4 \tau_5 \rangle$ = 1). The sites labeled $A,B,C$ identify with one another.}
	\label{fig:highT1}
\end{figure}

For the time being, we only consider twisted boundary conditions and
the corresponding gauge choice, since the analysis for the straight
case is extremely similar.  On the cylinder, the high temperature
expansion is very similar to the expansion in 2d, with the exception
that in addition to ``trivial'' loops, which already occur in 2d,
there are ``non-trivial'' loops, which wind around the cylinder.  It
is clear that the former, as in 2d, contribute always in a
translationally invariant manner. Let us consider the non-trivial
loops.  The leading such contribution for small $x$ is the one of
minimal length.  In the numerator, this must contain the bond $(mn)$,
as shown in Fig.~\ref{fig:highT1}, where the sublattices of the
magnetic unit cell is labeled $1,..,6$.   

To look for translational symmetry breaking, we investigate the
ratio of the correlation functions, e.g.
\begin{equation}
	\label{eq:highTcorr2}
	\displaystyle
	\frac{\langle J_{12} \tau_1(\mathbf{r}) \tau_2(\mathbf{r}) \rangle}{\langle J_{45} \tau_4(\mathbf{r}) \tau_5(\mathbf{r}) \rangle},
\end{equation}
which requires only the analysis of the numerator.  This is sufficient
to probe the translational symmetry breaking of the model for the case
of odd $L_y$.

However, because this lattice is composed of parallelograms, it is
always possible to form {\sl two} paths to move from one (red)
circular site to another.  These paths are {\sl not} equivalent and
are of the opposite sign because of the condition $\prod_a J_{ab} =
-1$ on each parallelogram.  This is depicted in Fig.~\ref{fig:highT1},
hich shows the two possible paths contributing to the correlation
$\langle \tau_4(\mathbf{r}) \tau_5(\mathbf{r}) \rangle$, both of the
order $x^5$ on $L_y = 3$ system.  These two paths have the opposite
sign and vanish under the sum in Eq.~\eqref{eq:highTcorr}.   

This demonstrates that the leading (in $x$) contributions to the
dimerization vanish for the nearest-neighbor model.  Similar
cancellations occur at higher orders.  For example, any longer loop
which encircles the cylinder and does not include compact closed
sub-loops always occurs with a partner of opposite sign.  In particular,
any path containing exactly two sides of any parallelogram will suffer
such a cancellation, as these two sides can be exchanged with the
opposite two, resulting in a new path of opposite sign.  The vast
majority of paths encircling the cylinder are of this type.  In fact, we
believe that the cancellation, and consequent absence of dimerization,
persists to all orders in $x$, and is an exact result.  However, we have
not proven this, owing to the complication of giving a general argument
which also includes paths which are ``decorated'' with many small closed
loops.  However, numerical analysis of finite systems supports our claim.

In summary, we expect that a nearest neighbor model does not break translational
symmetry, i.e. the ratio
\begin{equation}
	\displaystyle
	\frac{\langle J_{12} \tau_1(\mathbf{r}) \tau_2(\mathbf{r}) \rangle}{\langle J_{45} \tau_4(\mathbf{r}) \tau_5(\mathbf{r}) \rangle} = 1.
\end{equation}
\begin{figure}[t]
	\begin{center}
	\scalebox{0.32}{\includegraphics[]{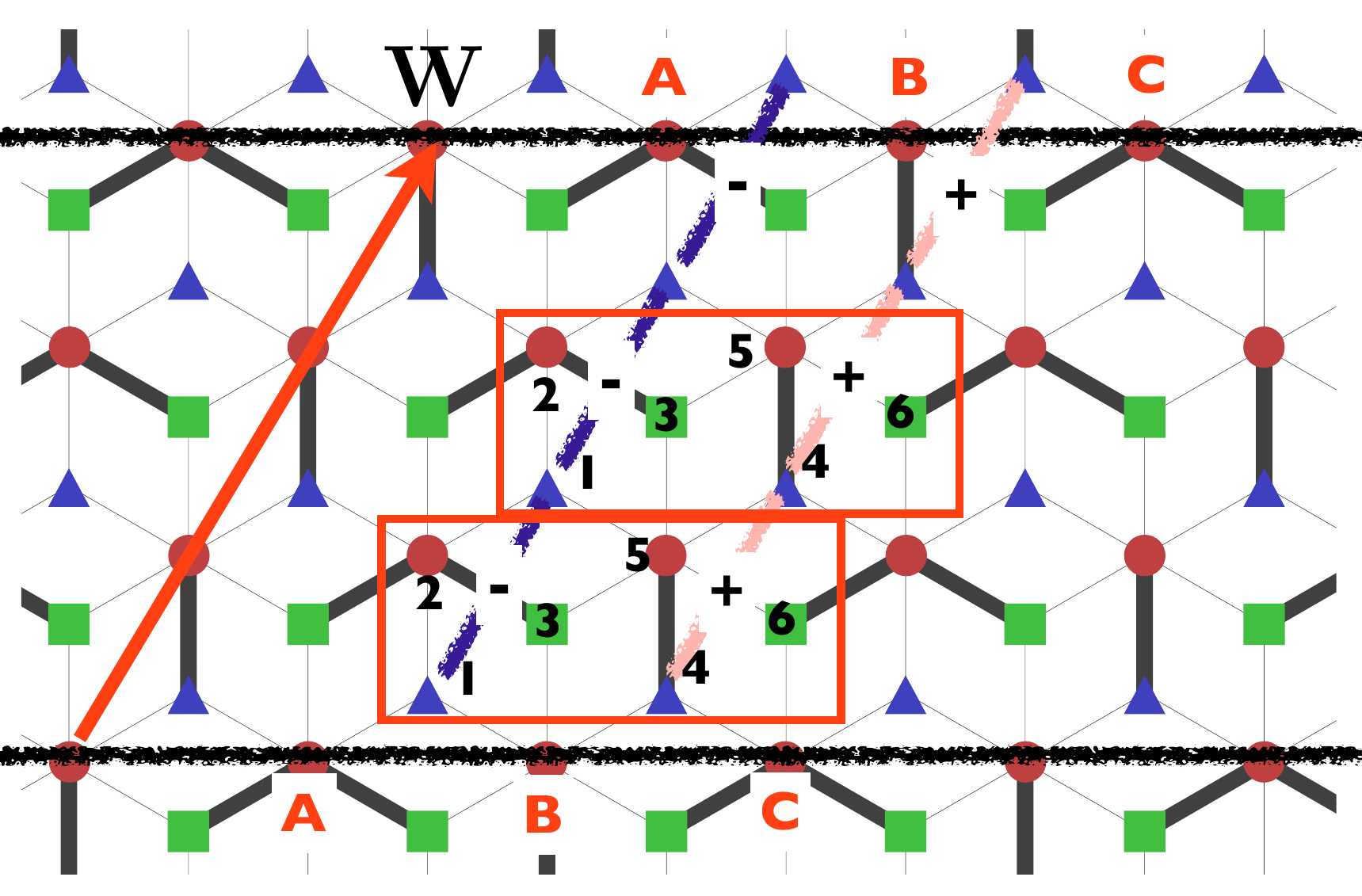}}
	\end{center}
	\caption{For $J_2/J_1 \gg 1$, there is one unique path, in the lowest order in $x$, that connects the (blue) triangular sites together. Notice that in odd $L_y$, it breaks translational symmetry. These two paths are shown in dark (blue) and light (red) lines.}
	\label{fig:highT2}
\end{figure}

The absence of dimerization is an accidental effect owing to the
destructive interference in the special geometry of the dice lattice.
We expect that further neighbor interactions will restore the expected
behavior.  Guided by the high temperature expansion, we seek to
introduce interactions which give a {\sl unique} path contributing to
leading order in $x$ for non-trivial loops around the cylinder, and
hence do not suffer destructive interference.  This is accomplished, for
instance, by allowing second neighbor interactions that couple the same
``type" of three-coordinated sites together.  In another words, the
triangular (blue) sites interact with each other and are independent of
the square (green) sites, which themselves interact with one another.
If we allow these signs to be staggered, we can consider the
correlations, to the lowest order in $x$, for odd $L_y$, (paths shown in
Fig.~\ref{fig:highT2}),
\begin{equation}
	\label{eq:highTcorr3}
	\frac{\langle J_{11}^{(2)} \tau_1(\mathbf{r}+\mathbf{W}) \tau_1(\mathbf{r}) \rangle}{\langle J_{44}^{(2)} \tau_4(\mathbf{r}+\mathbf{W}) \tau_4(\mathbf{r}) \rangle}.
\end{equation}
One can see graphically, from Fig.~\ref{fig:highT2}, that there is only one path, in the lowest order in $x$, that connects sites 1 and 1 offset by the vector $\mathbf{W}$.
Given the second neighbor couplings, schematically shown with ``+" and ``-" in Fig.~\ref{fig:highT2}, cylinders with an odd circumference must break translational symmetry.
If these interactions are allowed by symmetry, we can achieve a spontaneous translational symmetry breaking, where for odd leg systems, we expect the correlation function in Eq.~\eqref{eq:highTcorr3} is $-1$ in the lowest order in $x$.

\subsection{PSG analysis}
\label{sec:psg}

As mentioned above, we need further neighbor interactions to achieve
non-zero dimerization.  However, if we add in arbitrary second neighbor
interactions, we can, by hand, trivially break translational symmetry in
the bulk.  Our goal is to add further neighbor interactions which do not
{\sl explicitly} break any lattice symmetries, i.e. which in two
dimensions would preserve all physical symmetries of the system.  In the
dual Ising Hamiltonian, however, these symmetries are not manifest, due
to the fact that we had to make a gauge choice of the exchange
interactions $J_{ab}$.  The specific choice $J_{ab}$ clearly does not
preserve all lattice symmetries, and indeed there is no choice which
does.

As a consequence, a lattice symmetry operation must be accompanied by a
gauge transformation to preserve the form of $J_{ab}$.  The combined
operations of this type form the  projective symmetry group (PSG).  We
must work out the PSG, and then carefully
select further neighbor interactions that respect it. 

To formulate the PSG in more technical terms, we postulate that under a
lattice symmetry operation, $\mathcal{O}$, the Ising spins transform
according to
\begin{equation}
\mathcal{O} \tau_a^z \to s_{a'}\tau_{a'}^z 
\end{equation}
where $s_{a'} = \pm 1$ is the Ising gauge part of the operation, and $a
\rightarrow a'$ describes the lattice transformation.
Then, in the Hamiltonian in Eq.~\eqref{eq:dicehami},
\begin{equation}
J_{ab}  \tau_a^z \tau_b^z \to J_{ab} s_{a'} s_{b'} \tau_{a'}^z
\tau_{b'}^z \equiv J'_{a'b'} \tau_{a'}^z
\tau_{b'}^z,
\end{equation}
where $J'_{a'b'} = J_{ab} s_{a'} s_{b'}$.  For the Hamiltonian to be
invariant, therefore, the interactions must satisfy $J'_{a'b'} =
J_{a'b'}$, hence 
\begin{equation}
\label{eq:spinflip}
J_{ab} = J_{a'b'} s_{a'} s_{b'}.
\end{equation}
Given the nearest-neighbor pattern of interactions shown in
Fig.~\ref{fig:gauge}, Eq.~\eqref{eq:spinflip}  determines the pattern of spin flips $s_{a'}$ needed to
specify the PSG, for a given gauge choice and a given lattice symmetry
operation.  Details of the calculation of these spin flips is given in
Appendix~\ref{ap:psg1}. 

\begin{figure}[t]
	\begin{center}
	\scalebox{0.32}{\includegraphics[]{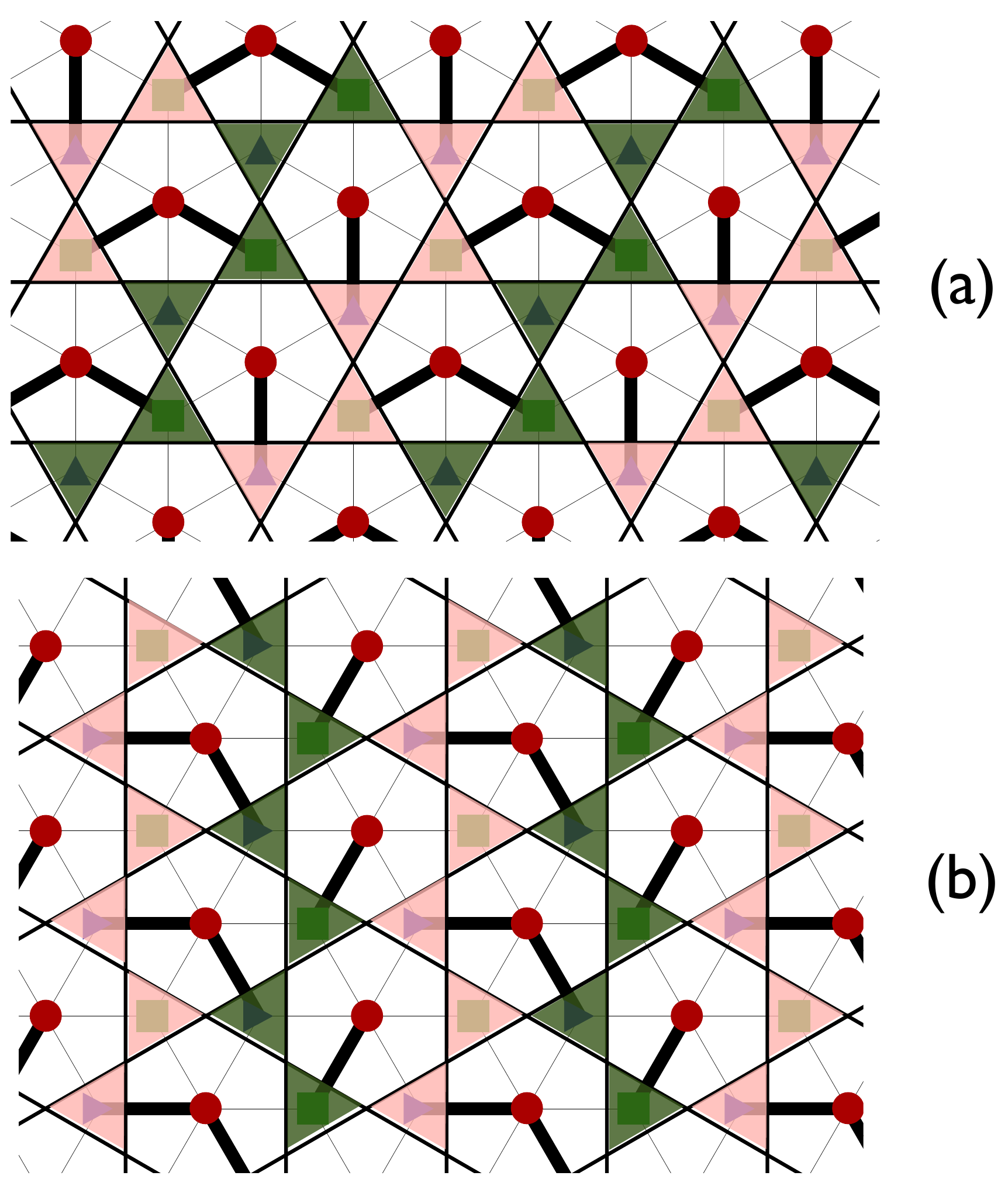}}
	\end{center}
	\caption{Sum of spin-spin correlations on each triangle for (a) twisted and (b) straight boundary conditions for cylinders with odd circumference.
	Here, the correlations on the dark (green) triangle are larger than that of the light (red) triangle.}
	\label{fig:sumtri}
\end{figure}

Then, we require that additional further neighbor interactions also
satisfy Eq.~\eqref{eq:spinflip} with {\em the same} spin flips $s_a$.
Out of the five possible further neighbor interactions up to lattice
spacing 2, there are only two symmetry allowed interactions consistent
with the PSG.  We consider only one of them here, in which an interaction
of the same strength connects the same three-coordinated lattice sites
(triangular (blue) to triangular).   With a proper choice of signs,
this interaction satisfies the PSG.  These signs are determined and
given in Appendix~\ref{ap:psg2}.

\subsection{Computing correlations}
\label{sec:corr}

\begin{figure}[t]
	\begin{center}
	\scalebox{0.32}{\includegraphics[]{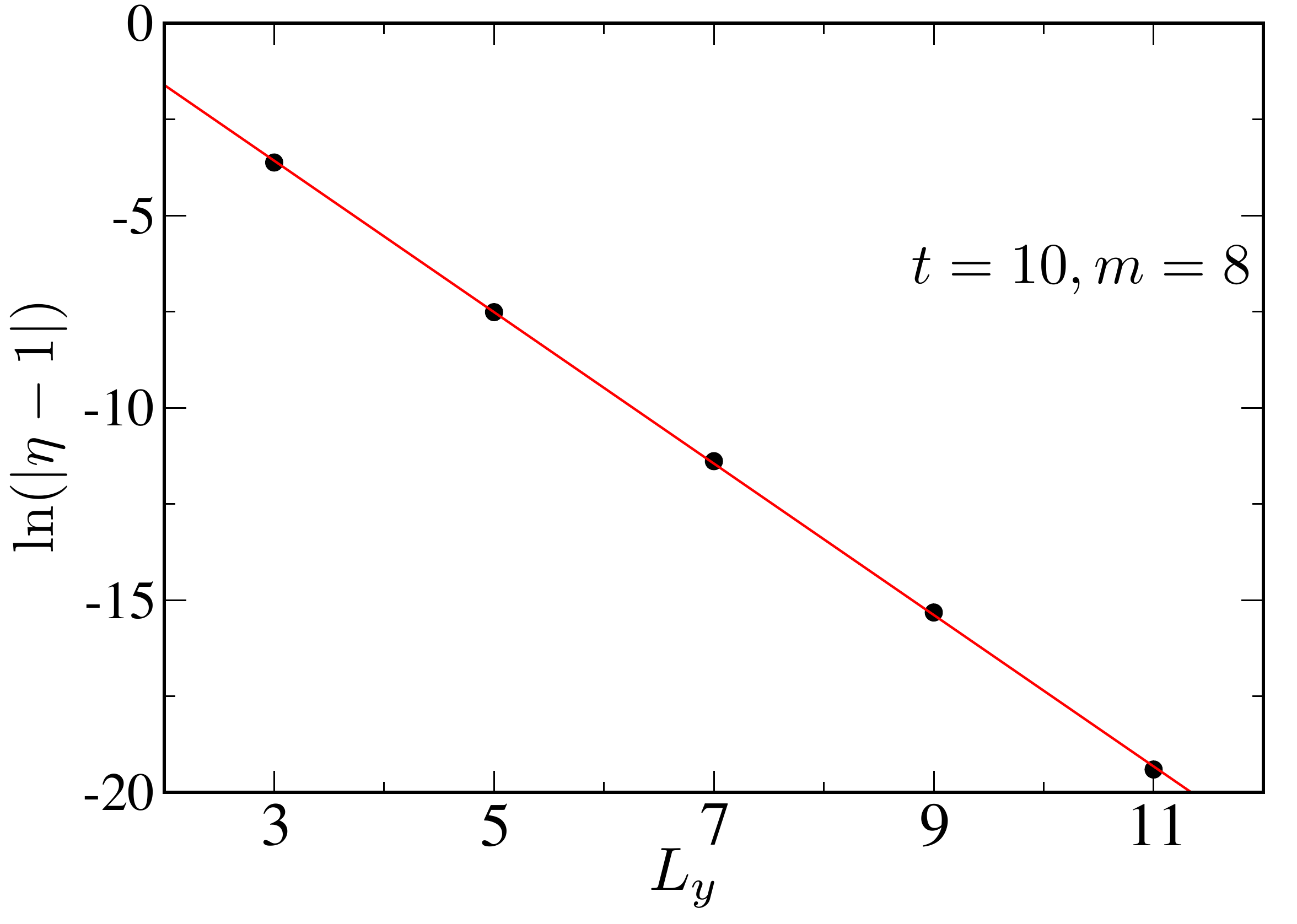}}
	\end{center}
	\caption{Here, we show the ratios of the (green) dark to (red) light triangles in Fig.~\ref{fig:sumtri}. The ratio decays exponentially as a function of $L_y$, meaning that in the 2d limit, we recover the $Z_2$ QSL, and the correlations throughout the lattice become uniform. The solid (red) line is a linear fit that yields a slope of $1.9(7)$.}
	\label{fig:ratio}
\end{figure}

We are now in a position to compute the dimerization operator.
Standard Landau reasoning implies that the universal parts of
quantities which have the same symmetry are proportional.  Hence, 
\begin{equation}
\label{eq:corr}
\la S_i S_j \lr = \la D_{ij} \lr \propto \la \sigma_{ij} \lr = \la J_{ab} \tau^z_a \tau^z_b \lr
\end{equation}
In the following discussions, we work in a path integral formulation
in the $\tau^z_a$ basis, where $\tau^z_a \to \phi_a$ is now softened
to take continuous values.
The action that corresponds to Eq.~\eqref{eq:dicehami} is
\begin{equation}
\label{eq:diceaction}
	S = \sum_{\omega,q} \phi^{(a)}_{\omega,q} \left[ (\omega^2 + m^2) I - J_q \right] \phi^{(b)}_{-\omega,-q}.
\end{equation}
Here, the sum over $a,b = 1,...,6$ is implicit, where the site indices are labeled in Fig.~\ref{fig:duality}.
$J_q$ is the Fourier transform of the interaction matrix $J_{ab}$, which is shown explicitly in Appendix~\ref{ap:jq} for both boundary conditions.
We have introduced a Matsubara frequency $\omega$ as well as a mass $m$.
Using this formalism, we are now in a position to compute the correlation function
\begin{equation}
\label{eq:maincorr}
\la J_{ab} \, \phi^{(a)} \phi^{(b)} \lr,
\end{equation}
by using the methods outlined in Appendix~\ref{ap:gfunc}.

We compare to the results of Ref.~\onlinecite{yan2011spin} by summing the spin-spin correlations for each triangle for both the boundary conditions.
For the case of even $L_y$, we obtain uniform correlations throughout the lattice: for instance, for the twisted boundary conditions, we cannot distinguish the triangle encompassing site 1 from that of site 4.
However, for the case of odd $L_y$, we obtain similar patterns to the DMRG results where there is some stripe order distinguishing the even $x$ from the odd $x$, which ultimately translates to a dimerization pattern that spontaneously break translational symmetry.
This is shown in Fig.~\ref{fig:sumtri}, where we graphically indicate the sum of correlations on each triangle on systems with odd circumferences.
To compute the correlations, we have set the mass $m=8$ and the relative strength of the second neighbor interactions as $t = 10$.
Additionally, we define a ratio of these correlations as
\begin{equation}
\eta \equiv \frac{\sum_{ab \in \text{even triangle}} \la J_{ab} \, \phi^{(a)} \phi^{(b)} \lr}{\sum_{ab \in \text{odd triangle}} \la J_{ab} \, \phi^{(a)} \phi^{(b)} \lr}
\end{equation}
and plot this as a function of odd $L_y$ in Fig.~\ref{fig:ratio}.
Notice that the ratio of the correlations decay exponentially in the circumference, such that we recover the $Z_2$ gauge theory in the two-dimensional limit.
In addition, given a set of parameters, $m, t$, both boundary conditions yield the same correlations and hence, the same $\eta$.
This is because both conditions have exactly the same second neighbor interactions, where the details are provided in Appendix~\ref{ap:psg2}.

\section{Conclusion}
\label{sec:conclude}

In this paper, we explored the even/odd effects of the $Z_2$ gauge
theory on the kagome lattice.  Using a dual formalism, we used arguments
from high temperature expansion as well as projective symmetry group
analysis to include further neighbor interactions to the dual
Hamiltonian.  Finally, we calculated and summed the correlation
functions on the triangles of the kagome lattice and obtain an ordering
pattern similar to results in Ref.~\onlinecite{yan2011spin}.  This
confirms the consistency of the dimerization observed in
Ref.~\onlinecite{yan2011spin} with a $Z_2$ quantum spin liquid state in the
thermodynamic limit.

After this work was completed, during the writing of this paper, we
became aware of related work by Wan and Tchernyshyov, which was recently
posted in Ref.~\onlinecite{2013arXiv1301.5008W}.  They use a different
$Z_2$ gauge theory, which is based on a quantum dimer model
approximation of the original kagome Heisenberg model, and obtain
similar results for the dimerization (as well as other results not
obtained here).  The approaches are complimentary, and the agreement
between dimerization patterns attests to the robustness of this finite
size phenomena in $Z_2$ spin liquid states.

\acknowledgements

We would like to thank Ru Chen and Yuan Wan for discussions.  This
research was supported by the DOE through BES grant DE-FG02-08ER46524,
and by Microsoft Station Q research center.  

\appendix

\section{High temperature expansion}
\label{ap:hight}

Here, we give a quick refresher of the high temperature expansion of the Ising model.
We start by simplifying the partition function in the following way (we drop the superscript $z$ for clarity)
\begin{eqnarray}
	&&Z  =  \sum_{\tau_k = \pm 1} \; e^{\beta h \sum_{\langle ab \rangle} J_{ab} \tau_a \tau_b} 
	      =  \sum_{\tau_k = \pm 1} \; \prod_{\langle ab \rangle} e^{\beta h J_{ab} \tau_a \tau_b} \nonumber\\
	     & = & \sum_{\tau_k = \pm 1} \; \prod_{\langle ab \rangle} \; \sum_{n = 0}^\infty \frac{\left( \beta h J_{ab} \tau_a \tau_b\right)^n}{n!}\\
	    & = & \sum_{\tau_k = \pm 1}  \prod_{\langle ab \rangle} \left[  \sum_{n = 0}^\infty \frac{\left( \beta h J_{ab} \tau_a \tau_b\right)^{2n}}{(2n)!} 
	    																+ \sum_{n = 0}^\infty \frac{\left( \beta h J_{ab} \tau_a \tau_b\right)^{2n+1}}{(2n+1)!} \right]\nonumber
\end{eqnarray}
Since $(J_{ab} \tau_a \tau_b)^{2n} = 1$ for all $n \in Z^+$ ($\tau_a^2 = 1, J_{ab}^2 = 1$),
\begin{eqnarray}
\label{eq:hightpart}
	Z & = & \sum_{\tau_k = \pm 1} \; \prod_{\langle ab \rangle} \; \left[ \cosh(\beta h) + \sinh(\beta h) J_{ab} \tau_a \tau_b \right ]\\
	    & = &   \cosh(\beta h) \sum_{\tau_k = \pm 1} \; \prod_{\langle ab \rangle} \; \left[ 1 + \tanh(\beta h) J_{ab} \tau_a \tau_b \right] \nonumber\\
   	   & = & \cosh(\beta h) \sum_{\tau_k = \pm 1} \; \prod_{\langle ab \rangle} \; \left[ 1 + x J_{ab} \tau_a \tau_b \right] \nonumber,
\end{eqnarray}
where $x = \tanh(\beta h )$.

\section{Projective symmetry group}

In this appendix, we give details of the projective symmetry group (PSG) analysis done for the model in Eq.~\eqref{eq:dicehami}.
Recall that this is only necessary to work out the second neighboring interactions, which was required by the heuristic argument given in Sec.~\ref{sec:hight}.
We first start by giving the independent symmetries of the dice lattice.
Then, we move onto discussing the procedures to obtain the PSG for the nearest neighbor model and conclude with a brief discussion about which second nearest neighbors are allowed within our theory.

%
\begin{figure}[t]
  \begin{center}
  \scalebox{0.7}{\includegraphics[width=\columnwidth]{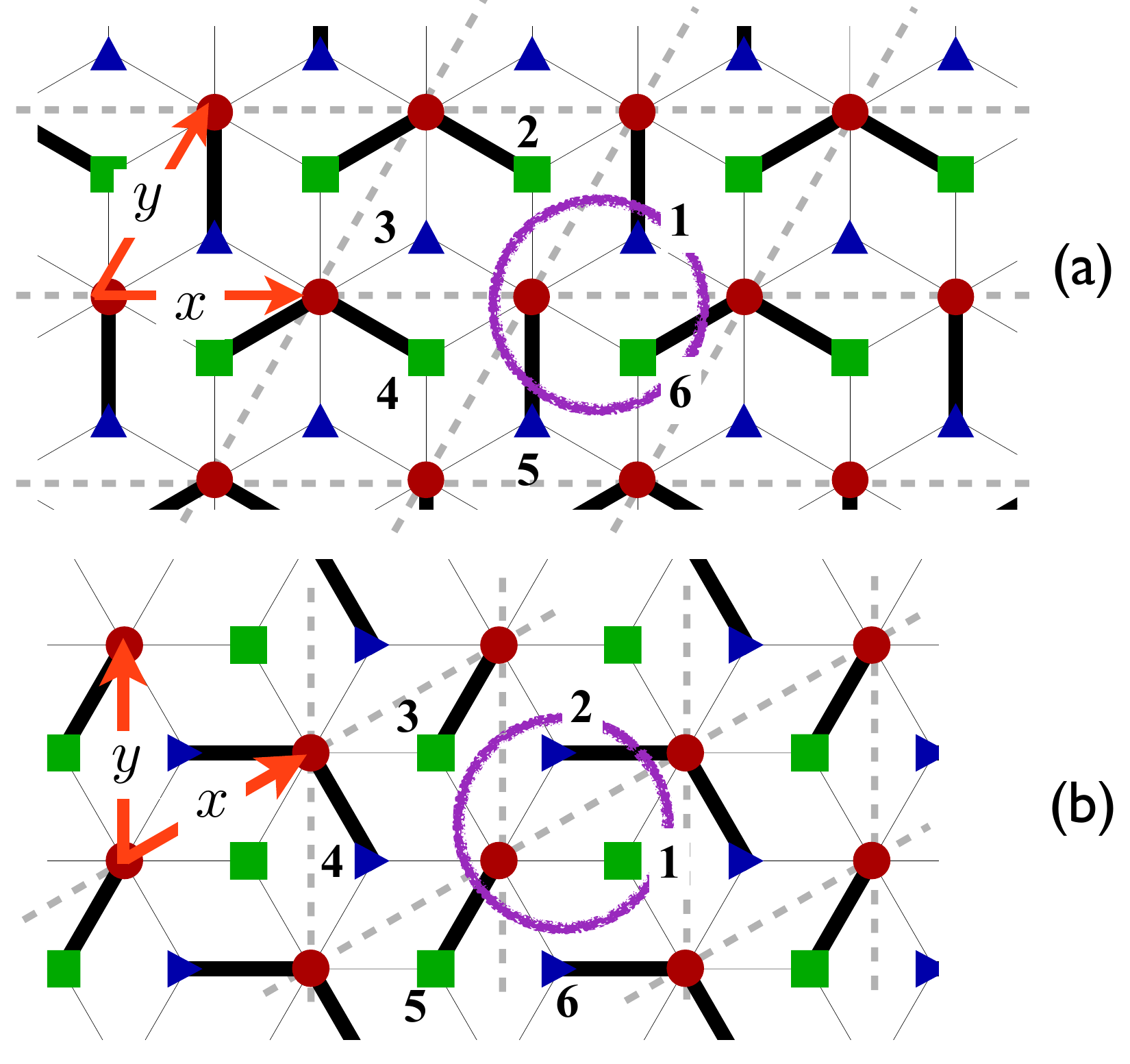}}
  \end{center}
  \caption{The basis directions are shown in (red) arrows and are labeled $x,y$.
		The faded dotted lines are the axes of each gauge choice.
		The sites that surround the (red) circular sites are labeled $1,...,6$.
		The (purple) open circle that enclose the three inequivalent sites is the unit cell for the dice lattice that have coordinates $(m,n)$ in their own respective bases.
		As before, we show the frustrated bonds, i.e. where $J_{ab} = -1$, in thick (black) lines while the unfrustrated bonds, $J_{ab} = 1$ in thin lines.
		Notice that the two geometries (a,b) are at $90^o$s from each other and have the same initial gauge description, given in Eq.~\eqref{eq:1}.}
        \label{fig:equiv}
\end{figure}
%

\subsection{Symmetries of the dice lattice}
\label{ap:equiv}

In the main text, we mentioned that there were three reflection symmetries on the dice lattice about the circular (red) site.
While this is true, these three reflection symmetries are not all independent -- namely, they can be generated using rotations and a single reflection symmetry.
Consider Fig.~\ref{fig:duality}.
There is reflection symmetry through the vertical line going through the circular site, which we will call $P_y$, through an upward sloping line, $P_+$, and through a downward sloping line, $P_-$.

\begin{eqnarray}
\label{eq:reflect}
 I & = & R_{\pi/3}^2 P_+ P_y, \\
 I & = & R_{\pi/3}^{-2} P_- P_y, \nonumber
\end{eqnarray}
where $I$ is the identity operator, and $R_\theta^{-1} = R_{-\theta}$.
Using these identities, we can conclude that the dice lattice is invariant under 2 point symmetries, rotation and reflection about a single axis, as well as 2 translations, labeled $\mathbf{u},\mathbf{v}$ in Fig.~\ref{fig:duality}.

\subsection{Nearest neighbor PSG}
\label{ap:psg1}

The two gauge choices have different winding vectors, where one ``twists" the original lattice while the other ``straight" vector does not.
The shorter, winding direction is denoted by $\mathbf{W}$, and the infinite direction by $\mathbf{a}_1$.
Let $(m,n)$ denote the coordinates of the circular (red) site: $m \hat{x} + n \hat{y}$.
We label the 6 neighbors of each circular (red) site from $1, ..., 6$ as shown in Fig.~\ref{fig:equiv}, and set the circular site as the origin.
Then, the gauge description, $J_{ab}$, can be written as $J_{mn}^\alpha$, where $\alpha = 1, ..., 6$ labels the 6 neighbors and $(m,n)$ describe the coordinates of the center circular (red) site.
Again, taking the thick lines as $J_{ab} = -1$ and the thin lines as $J_{ab} = 1$, we can describe both of these gauge choices as follows
\begin{eqnarray}
\label{eq:1}
  J_{mn}^1 & = & J_{mn}^2 = J_{mn}^3 = 1 \\
  J_{mn}^4 & = & J_{mn}^6 = (-1)^m \nonumber \\
  J_{mn}^5 & = & -(-1)^m \nonumber.
\end{eqnarray}
The idea of the PSG scheme is to first transform the lattice and gauge structure under lattice symmetries.
Then work out the pattern of spin flips, using Eqs. (\ref{eq:spinflip},\ref{eq:1}), that restores our original gauge choices.

\begin{table}[t]
\begin{tabular}{cl||c|c|c|cc}
\centering
& Original			& $T_y$ 		& $T_x$ 			& $R_{\pi/3}$		& $P_{25}$	\\ \toprule 
& (123456) 		& (123456) 	& (123456)		& (612345)		& (321654)	\\ \hline
& $(m,n)$		 	& $(m,n+1)$	& $(m+1,n)$		& $(m+n,-m)$ 		& $(-m-n,n)$	\\ \hline
\end{tabular}
\caption{Transformation of sites and coordinates under lattice operations for the cylinder with {\sl twisted} boundary conditions.
		Here, $P_{25}$ is reflection symmetry about the axes connecting sites 2 and 5.}
\label{tab:twistedtransform} 
\begin{tabular}{cl||c|c|c|cc}
\centering
& Original			& $T_y$ 		& $T_x$ 			& $R_{\pi/3}$		& $P_{25}$	\\ \toprule 
& (123456) 		& (123456) 	& (123456)		& (612345)		& (321654)	\\ \hline
& $(m,n)$		 	& $(m,n+1)$	& $(m+1,n)$		& $(m+n,-m)$ 		& $(n,m)$		\\ \hline
\end{tabular}
\caption{Transformation of sites and coordinates under lattice operations for the cylinder with {\sl straight} boundary conditions.
		Here, $P_{25}$ is reflection symmetry about the axes connecting sites 2 and 5.}
\label{tab:straighttransform} 
\end{table}

We first observe how the site indices and coordinates $(m,n)$, in Fig.~\ref{fig:equiv}, change under lattice transformations.
As mentioned in the previous subsection, we consider two translations ($x$, $y$ in their respective gauge choices), a single $\pi/3$ rotation and a reflection about the axis connecting sites 2 and 5, which we denote as $P_{25}$.
Tables \ref{tab:twistedtransform},\ref{tab:straighttransform} show how the site indices as well as the positions of these sites change under these transformations.
The first row shows how the site indices permute in case of a lattice transformation.
Take for instance, $R_{\pi/3}$, which is $\pi/3$ rotations about the (red) circular site.
From Fig.~\ref{fig:equiv}, it is clear that site 6 takes the place of site 1, 1 of 2, etc.
This is tabulated by $(123456) \to (612345)$.
The second row shows how the position of the unit cell changes under these operations.
For instance, translations under $y$ takes the coordinate $(m,n) \to (m, n+1)$.

Next, we solve for $s_{a'} s_{b'}$ using Eq.~\eqref{eq:spinflip}, where the original $J_{ab}$ is given by Eq.~\eqref{eq:1}.
These are tabulated in Table~\ref{tab:twisted1st} for the twisted boundary condition and in Table~\ref{tab:straight1st} for the straight boundary condition.
For each spin flip, $s_a = \pm1$, we label its position $(m,n)$ using the unit cell in Fig.~\ref{fig:equiv}, and the type $(p,b,g)$ for the (red, blue, green) or (circle, triangle, square) sites respectively.
\begin{table}[t]
\begin{tabular}{cl|c||c|c|c|c}
\centering
&Site	& Spin flips					& $T_y$ 	& $T_x$ 	& $R_{\pi/3}$	& $P_{25}$	\\ \toprule 
&1		& $s_{mn}^p s_{mn}^b$ 		& 1 		& 1 		& $(-1)^{m+n}$	& 1			\\ \hline
&2		& $s_{mn}^p s_{m-1,n+1}^g$ 	& 1 		& 1	 	& 1 			& 1			\\ \hline
&3		& $s_{mn}^p s_{m-1,n}^b$ 	& 1 		& 1	 	& 1 			& 1			\\ \hline
&4		& $s_{mn}^p s_{m-1,n}^g$ 	& 1 		& -1	 	& $(-1)^m$	& $(-1)^n$	\\ \hline
&5		& $s_{mn}^p s_{m,n-1}^b$ 	& 1 		& -1	 	& $-(-1)^n$	& $(-1)^n$	\\ \hline
&6		& $s_{mn}^p s_{mn}^g$ 		& 1 		& -1	 	& $-(-1)^n$	& $(-1)^n$	\\ \hline
\end{tabular}
\caption{PSG for nearest neighbor interactions on the cylinder with {\sl twisted} boundary conditions.}
\label{tab:twisted1st} 
\begin{tabular}{cl|c||c|c|c|c}
\centering
&Site	& Spin flips					& $T_y$ 	& $T_x$ 	& $R_{\pi/3}$	& $P_{25}$		\\ \toprule 
&1		& $s_{mn}^p s_{mn}^g$ 		& 1 		& 1	 	& $(-1)^{m+n}$	& 1				\\ \hline
&2		& $s_{mn}^p s_{mn}^b$ 		& 1 		& 1 		& 1			& 1				\\ \hline
&3		& $s_{mn}^p s_{m-1,n+1}^g$ 	& 1 		& 1	 	& 1 			& 1				\\ \hline
&4		& $s_{mn}^p s_{m-1,n}^b$ 	& 1 		& -1	 	& $(-1)^m$	& $(-1)^{m+n}$		\\ \hline
&5		& $s_{mn}^p s_{m-1,n}^g$ 	& 1 		& -1	 	& $-(-1)^n$	& $(-1)^{m+n}$		\\ \hline
&6		& $s_{mn}^p s_{m,n-1}^b$ 	& 1 		& -1	 	& $-(-1)^n$	& $(-1)^{m+n}$		\\ \hline
\end{tabular}
\caption{PSG for nearest neighbor interactions on the cylinder with {\sl straight} boundary conditions.}
\label{tab:straight1st} 
\end{table}
Take, for instance, translations along $x$, where the coordinates of each site changes to $(m+1,n)$.
Given Eq.~\eqref{eq:1}, because the $J_{ab}$ on sites $1,2,3$ do not depend on the coordinate $m$, the product of spin flips on the (red) circular site and sites $1,2,3$ are equal to 1.
However, the $J_{ab}$ on sites $4,5,6$ are dependent on $m$ and the product of the spin flips, since $J_{m+1,n} = -J_{m,n}$ for each of these sites, is equal to -1.
On the other hand, because the $J_{ab}$ do not depend on coordinate $n$, translation in the $y$-direction does not affect the gauge structure.
The spin flip patterns for the other two point symmetries are computed in a similar fashion.
While it may be possible to solve for each $s_{m,n}$, it turns out that it is unnecessary to do so because to compute the next nearest interactions, we use the fact that $s_{mn}^2 = 1$ for all $(p,b,g)$ and coordinates.

\subsection{Including second nearest neighbor interactions}
\label{ap:psg2}

\begin{table}[t]
\begin{tabular}{cl|c|c|c|c|c}
\centering
& $a$ 	& $b$ 	& $c$	& $d$	&$e$	&$f$		\\ \hline 
& 1-3	& 2-4	& 3-5	& 4-6	& 5-1	& 6-2	
\end{tabular}
\caption{Connections for the 2nd neighbor interactions.
		For instance, $a$ corresponds to the interaction between sites 1 and 3.}
\label{tab:2ndinteract} 
\end{table}

We consider further neighbors up to lattice spacing 2.
There are a total of 5 type of interactions, of which only 2 are allowed by PSG.
We consider one of these interactions, which connects the {\sl same} three-coordinated lattice sites (e.g. sites 1 and 3 in Fig.~\ref{fig:equiv}).
The relative strength of the interaction with respect to the nearest-neighbor coupling is denoted by $t$; however, the relative signs of these interactions must be chosen to be consistent with the PSG.
The connections of this interaction are shown in Table~\ref{tab:2ndinteract}.
These signs will be labeled as $\alpha_{mn}$, where $\alpha = a,b,c,d,e,f$ and $m,n$ denotes the position of the red site.
Here, we take the form $\alpha_{mn} = (-1)^{g_\alpha(m,n)}$ and require that the strength of the couplings are equal, i.e.
\begin{equation}
J_{mn}^{\text{2nd},\alpha}/J = t \, \alpha_{mn}.
\end{equation}
We can use the PSG equations from Sec.~\ref{ap:psg1} to compute $g_\alpha(m,n)$ for each interaction $\alpha$.

\begin{table}[t]
\begin{tabular}{cl||c|c|c|cc}
\centering
& Original			& $T_y$ 		& $T_x$ 			& $R_{\pi/3}$		& $P_{25}$	\\ \toprule 
& (abcdef) 		& (abcdef) 	& (abcdef)			& (fabcde)			& (afedcb)		\\ \hline
& $(m,n)$		 	& $(m,n+1)$	& $(m+1,n)$		& $(m+n,-m)$ 		& $(-m-n,n)$	\\ \hline
\end{tabular}
\caption{PSG for nearest neighbor interactions on the cylinder with {\sl twisted} boundary conditions.}
\label{tab:twistedtransform2} 
\begin{tabular}{cl||c|c|c|cc}
\centering
& Original			& $T_y$ 		& $T_x$ 			& $R_{\pi/3}$		& $P_{25}$	\\ \toprule 
& (abcdef) 		& (abcdef) 	& (abcdef)			& (fabcde)			& (afedcb)		\\ \hline
& $(m,n)$		 	& $(m,n+1)$	& $(m+1,n)$		& $(m+n,-m)$ 		& $(n,m)$		\\ \hline
\end{tabular}
\caption{PSG for nearest neighbor interactions on the cylinder with {\sl straight} boundary conditions.}
\label{tab:straighttransform2} 
\end{table}

We now give a detailed description of our procedures.
Because both boundary conditions follow the same transformations under translations, we consider one of them here.
Under $y$-translation, the interactions $(abcdef) \to (abcdef)$ and the coordinates change from $(m,n) \to (m,n+1)$.
Let us concentrate for example, on the $a$ interaction, which is between (blue) triangular sites located at $(m,n)$ and $(m-1,n)$.
We use Eq.~\eqref{eq:spinflip}, where the original $J_{ab} = a_{mn}$ and the transformed $J_{a'b'} = a_{m,n+1}$.
The transformations of each of these interactions under various lattice symmetries are tabulated in Tables (\ref{tab:twistedtransform2},\ref{tab:straighttransform2}).
Then, since $s_{mn}^2 = 1$,
\begin{eqnarray}
J_{ab}/J_{a'b'} = a_{mn}/ a_{m,n+1} & = & s_{mn}^b s_{m-1,n}^b   \\
							& = & s_{mn}^b (s_{mn}^p)^2 s_{m-1,n}^b \nonumber \\
							& = & (s_{mn}^b s_{mn}^p) \, (s_{m-1,n}^b s_{mn}^p ) \nonumber \\
							& = & 1 \times 1 = 1 \nonumber
\end{eqnarray}

In going from the third to the fourth line, we used the nearest neighbor PSG in Tables (\ref{tab:twisted1st},\ref{tab:straight1st}) to get the products in the parentheses in the third line.
Using similar techniques, we can solve for the ratios of interactions $(abcdef)$ and their respective $g_\alpha(m,n)$.
These are tabulated in Tables (\ref{tab:twisted2},\ref{tab:straight2}).
There, since the ratios of the interactions are equal to 1 under $y$-translation, the interactions, $\alpha$, do not have any $n$ dependence, i.e. $g_\alpha(m,n) = g_\alpha(m)$.
Using the $x$-translations, however, one can obtain either $g_\alpha(m) = 2$ or $m$.
Then, using either rotations or reflection, solve for the relative signs between the 6 interactions, and make sure that it stays consistent with the nearest neighbor PSG.
It turns out that both of these boundary conditions produce the same results of $(abcdef)$ interactions.
\begin{eqnarray}
a_{mn} & = & d_{mn} = 1 \\
b_{mn} & = & f_{mn} = (-1)^m \\
c_{mn} & = & e_{mn} = -(-1)^m
\end{eqnarray}
where we have chosen the sign $a_{mn} = +1$ (since we can only get the relative signs of each interaction through this analysis).
Notice again that these signs are only dependent on the coordinate $m$, which is expected since the original gauge choices in Eq.~\eqref{eq:1} are only dependent on $m$ as well.

\begin{widetext}

\begin{table}[h]
\begin{tabular}{c|c||c|c|c|c}
\centering
Interaction	&Spin flips					& $T_y$ 					& $T_x$ 					& $R_{\pi/3}$						& $P_{25}$					\\ \toprule 
1-3			&$s_{mn}^b s_{m-1,n}^b$ 		& $a_{mn}/a_{m,n+1} = 1$	& $a_{mn}/a_{m+1,n} = 1$ 	& $a_{mn}/f_{m+n,-n} = (-1)^{m+n}$		& $a_{mn}/a_{-m-n,n} = 1$		\\ \hline
2-4			&$s_{m-1,n+1}^g s_{m-1,n}^g$ 	& $b_{mn}/b_{m,n+1} = 1$	& $b_{mn}/b_{m+1,n} = -1$	& $b_{mn}/a_{m+n,-n} = (-1)^m$		& $b_{mn}/f_{-m-n,n} = (-1)^n$		\\ \hline
3-5			&$s_{m-1,n}^b s_{m,n-1}^b$ 		& $c_{mn}/c_{m,n+1} = 1$ 	& $c_{mn}/c_{m+1,n} = -1$	& $c_{mn}/b_{m+n,-n} = -(-1)^n$		& $c_{mn}/e_{-m-n,n} = (-1)^n$		\\ \hline
4-6			&$s_{m-1,n}^g s_{m,n}^g$ 		& $d_{mn}/d_{m,n+1} = 1$	& $d_{mn}/d_{m+1,n} = 1$	& $d_{mn}/c_{m+n,-n} = -(-1)^{m+n}$	& $d_{mn}/d_{-m-n,n} = 1$		\\ \hline
5-1			&$s_{m,n-1}^b s_{m,n}^b$ 		& $e_{mn}/e_{m,n+1} = 1$	& $e_{mn}/e_{m+1,n} = -1$	& $e_{mn}/d_{m+n,-n} = -(-1)^m$		& $e_{mn}/c_{-m-n,n} = (-1)^n$		\\ \hline
6-2			& $s_{mn}^g s_{m-1,n+1}^g$ 		& $f_{mn}/f_{m,n+1} = 1$ 		& $f_{mn}/f_{m+1,n} = -1$		& $f_{mn}/e_{m+n,-n} = -(-1)^n$		& $f_{mn}/b_{-m-n,n} = (-1)^n$		\\ \hline
\end{tabular}
\caption{PSG for 2nd nearest neighbor interactions on the cylinder with {\sl twisted} boundary conditions.}
\label{tab:twisted2} 
\begin{tabular}{c|c||c|c|c|c}
\centering
Interaction	&Spin flips					& $T_y$ 					& $T_x$ 					& $R_{\pi/3}$						& $P_{25}$						\\ \toprule 
1-3			&$s_{mn}^g s_{m-1,n+1}^g$ 		& $a_{mn}/a_{m,n+1} = 1$ 	& $a_{mn}/a_{m+1,n} = 1$ 	&  $a_{mn}/f_{m+n,-n} = (-1)^{m+n}$		& $a_{mn}/a_{n,m} = 1$				\\ \hline
2-4			&$s_{mn}^b s_{m-1,n}^b$ 		& $b_{mn}/b_{m,n+1} = 1$ 	& $b_{mn}/b_{m+1,n} = -1$	&  $b_{mn}/a_{m+n,-n} = (-1)^{m}$		& $b_{mn}/f_{n,m} = (-1)^{m+n}$		\\ \hline
3-5			&$s_{m-1,n+1}^g s_{m-1,n}^g$	& $c_{mn}/c_{m,n+1} = 1$ 	& $c_{mn}/c_{m+1,n} = -1$	&  $c_{mn}/b_{m+n,-n} = -(-1)^n$ 		& $c_{mn}/e_{n,m} = (-1)^{m+n}$		\\ \hline
4-6			&$s_{m-1,n}^b s_{m,n-1}^b$ 		& $d_{mn}/d_{m,n+1} = 1$ 	& $d_{mn}/d_{m+1,n} = 1$	&  $d_{mn}/c_{m+n,-n} = -(-1)^{m+n}$	& $d_{mn}/d_{n,m} = 1$				\\ \hline
5-1			&$s_{m-1,n}^g s_{mn}^g$ 		& $e_{mn}/e_{m,n+1} = 1$ 	& $e_{mn}/e_{m+1,n} = -1$	&  $e_{mn}/d_{m+n,-n} = -(-1)^{m}$		& $e_{mn}/c_{n,m} = 1$				\\ \hline
6-2			&$s_{m,n-1}^b s_{mn}^b$ 		& $f_{mn}/f_{m,n+1} = 1$	 	& $f_{mn}/f_{m+1,n} = -1$	 	&  $f_{mn}/e_{m+n,-n} = -(-1)^n$		& $f_{mn}/b_{n,m} = (-1)^{m+n}$		\\ \hline
\end{tabular}
\caption{PSG for 2nd nearest neighbor interactions on the cylinder with {\sl straight} boundary conditions.}
\label{tab:straight2} 
\end{table}

\end{widetext}

\section{Computing correlation functions}
\label{ap:gfunc}

Here, we compute the Green's function from the action given in Eq.~\eqref{eq:diceaction}, which is
\begin{equation}
	\label{corr:1}
	S = \sum_{\omega,q} \phi_{\omega,q} G_{q,\omega}^{-1} \phi_{-\omega,-q},
\end{equation}
where $G_{q,\omega}^{-1} = (\omega^2 + m^2) I - J_q$.
At zero temperature, the sum over $\omega$ becomes an integral.
Then, the goal is to compute the Green's function to obtain the correlation functions.
We proceed by writing the Green's function in the following way
\begin{eqnarray}
	\label{corr:2}
	G_{q} = \int_\omega \left[ (\omega^2+m^2) I - J_q \right]^{-1},
\end{eqnarray}
where $G_q$ is a function of $q$.
We can diagonalize $J_q$ for a given $q$, i.e. $J_q | q \rangle = \epsilon_q | q \rangle$.
Notice that this cannot be analytically diagonalized for a general $J_q$, especially since the matrix is 6 by 6 in our case.
We can, however, numerically diagonalize the matrix $J_q$ as $U^{-1}_q J_q U_q = D_q$, where $U$ is a matrix of eigenvectors independent of $\omega$, and $D_q$ is the diagonal matrix consisting of eigenvalues that correspond to $U_q$.
Then, we can use this to integrate through $\omega$ (we omit the subscript $q$ for clarity)
\begin{eqnarray}
	\label{corr:3}
	U^{-1} G_{q} U	 & = & \int_\omega U^{-1} \left[ (\omega^2+m^2) I - J_q \right]^{-1} U\\
					 & = & \int_\omega \left[U^{-1} (\omega^2+m^2) I U - U^{-1} J_q U \right]^{-1}  \nonumber\\
					 & = & \int_\omega \left[ (\omega^2+m^2) I - \epsilon_q I \right]^{-1}  \nonumber\\
					 & = & \int_\omega \frac{1}{(\omega^2 + m^2) - \epsilon_q} I  \nonumber\\
					 & = & \left[ \frac{1}{2\sqrt{m^2 - \epsilon_q}} \right]\nonumber.
\end{eqnarray}
We can finally solve for the Green's function to get the correlation function in Eq.~\eqref{eq:corr}
\begin{eqnarray}
	\label{corr:4}
	G_{q}		& = & U \left[ \frac{1}{2\sqrt{m^2 - \epsilon_q}} \right] U^{-1},
\end{eqnarray}
where $G_q$ is a 6 by 6 matrix, where the indices label the sublattice, shown in Fig.~\ref{fig:gauge}.

\section{Interaction matrices}
\label{ap:jq}

Here, we show the Fourier transformed interaction matrices necessary to compute the correlation function in Eq.~\eqref{eq:maincorr}.
The first subsection shows that of the twisted boundary conditions while the second shows that of the straight boundary conditions.
Notice that including only the nearest neighbor interaction gives rise to three, doubly degenerate flat bands.
Including second neighbor interactions, discussed in Appendix~\ref{ap:psg2}, introduces curvature to these flat bands.

\subsection{Twisted boundary conditions}

In this section, we show the interaction matrices of the $Z_2$ gauge theory with twisted boundary conditions.
The indices $1, ..., 6$ are labeled in Fig.~\ref{fig:gauge}(a).
The interaction matrices are as follows:

\begin{widetext}
\[
J_q^{(1)} = 
\begin{pmatrix}
	\label{eq:jq}
	\displaystyle
	0 & -1 + \gamma_1 & 0 & 0 & \gamma_1 & 0 \\
	-1 + \gamma_1^* & 0 & 1 & \gamma_1^* \gamma_2 & 0 & \gamma_2 + \gamma_1^* \gamma_2\\
	0 & 1 & 0 & 0 & -1 + \gamma_1 & 0\\
	0 & \gamma_1 \gamma_2^* & 0 & 0 & 1 + \gamma_1 & 0\\
	\gamma_1^* & 0 & -1 + \gamma_1^* & 1+\gamma_1^*&  0 &-1\\
	0 & \gamma_2^* + \gamma_1 \gamma_2^* & 0 & 0 & -1 & 0
\end{pmatrix}, \quad
J_q^{(2)} = 
t
\begin{pmatrix}
	\displaystyle
	-\kappa_1 & 0 & 0 & \kappa_2 & 0 & 0 \\
	0 & 0 & 0 &0 & 0 & 0\\
	0 & 0 & -\kappa_1 & 0 & 0 & \kappa_3\\
	\kappa_2^* & 0 & 0 & \kappa_1 & 0 & 0\\
	0 & 0 & 0 & 0 &  0 & 0\\
	0 & 0 & \kappa_3^* & 0 & 0 & \kappa_1
\end{pmatrix},
\]
\end{widetext}
where 
\begin{eqnarray}
	\displaystyle
	\gamma_1 & = & e^{i \mathbf{q} \cdot \mathbf{W}} \nonumber \\
	\gamma_2 & = & e^{i \mathbf{q} \cdot \mathbf{a}_1} \nonumber \\
	\kappa_1 &=&  e^{-i \mathbf{q} \cdot \mathbf{W}} + e^{i \mathbf{q} \cdot \mathbf{W}} \nonumber\\
	\kappa_2 &=&  1 + e^{i \mathbf{q} \cdot \mathbf{a}_1} + e^{i \mathbf{q} \cdot \mathbf{W}} - e^{-i \mathbf{q} \cdot (\mathbf{W} - \mathbf{a}_1) } \nonumber\\
	\kappa_3 &=&  1 + e^{i \mathbf{q} \cdot \mathbf{a}_1} - e^{i \mathbf{q} \cdot \mathbf{W}} + e^{-i \mathbf{q} \cdot (\mathbf{W} - \mathbf{a}_1) }  \nonumber
\end{eqnarray}
and $t = J_2/J_1$.
Here, the vectors are shown in Fig.~\ref{fig:gauge}(a), where $\mathbf{W} = (\sqrt{3}/2,\sqrt{3})$ and $\mathbf{a}_1 = (2\sqrt{3},0)$.

\subsection{Straight boundary conditions}

In this section, we show the interaction matrices of the $Z_2$ gauge theory with twisted boundary conditions.
The indices $1, ..., 6$ are labeled in Fig.~\ref{fig:gauge}(b).
The nearest neighbor interactions are given as follows

\begin{widetext}
\[
J_q^{(1)} = 
\begin{pmatrix}
	\displaystyle
	0 & 0 & -1 & 0 & 0 & \gamma_2+ \gamma_1 \gamma_2 \\
	0 & 0 & 1+\gamma_1^* & 0 & 0 & \gamma_2 \\
	-1 & 1+\gamma_1 & 0 & 1-\gamma_1 & 1 & 0 \\
	0 & 0 & 1 -\gamma_1^* & 0 & 0 & 1 \\
	0 & 0 & 1 & 0 & 0 & -1 + \gamma_1 \\
	\gamma_2^* +\gamma_1^* \gamma_2^* & \gamma_2^* & 0 & 1 & -1 + \gamma_1^*& 0
\end{pmatrix}, \quad
J_q^{(2)} = 
t
\begin{pmatrix}
	\label{eq:jq}
	\displaystyle
	\kappa_6 & 0 & 0 & \kappa_4 & 0 & 0 \\
	0 & \kappa_6 & 0 & 0 & \kappa_5& 0 \\
	0 & 0 & 0 & 0 & 0 & 0 \\
	\kappa_4^* & 0 & 0 & -\kappa_6 & 0 & 0 \\
	0 & \kappa_5^* & 0 & 0 & -\kappa_6& 0 \\
	0 & 0 & 0 & 0 & 0 & 0
\end{pmatrix},
\]

\end{widetext}

where 
\begin{eqnarray}
	\displaystyle
	\kappa_4 &=&  -1 + e^{i \mathbf{q} \cdot \mathbf{W}} + e^{i \mathbf{q} \cdot \mathbf{a}_1} + e^{i \mathbf{q} \cdot (\mathbf{a}_1 + \mathbf{W})} \nonumber\\
	\kappa_5 &=&   1 + e^{-i \mathbf{q} \cdot \mathbf{W}} + e^{i \mathbf{q} \cdot (\mathbf{W}-\mathbf{a}_1)} - e^{i \mathbf{q} \cdot \mathbf{a}_1 } \nonumber\\
	\kappa_6 &=&  2 \cos ( \mathbf{q} \cdot \mathbf{W}).  \nonumber
\end{eqnarray}
Here, the vectors are shown in Fig.~\ref{fig:gauge}(b), where $\mathbf{W} = (0,\sqrt{3})$ and $\mathbf{a}_1 = (3,0)$.

\end{document}